# Three-dimensional porous graphene networks expand graphene-based electronic device applications

Yoshikazu Ito[a,b,†], Yoichi Tanabe[c], Katsuaki Sugawara[d,e], Mikito Koshino[f], Takashi Takahashi[c,d,e], Katsumi Tanigaki[c,d], Hideo Aoki[g,h,i] and Mingwei Chen[d,j]

[a]Institute of Applied Physics, Graduate School of Pure and Applied Sciences, University of Tsukuba, Tsukuba, Ibaraki 305-8573, Japan.

[b]PRESTO, Japan Science and Technology Agency, Saitama 332-0012, Japan.

[c]Department of Physics, Graduate School of Science, Tohoku University, Sendai, 980-8578, Japan.

[d]WPI Advanced Institute for Materials Research, Tohoku University, Sendai 980-8577, Japan.

[e]Center for Spintronics Research Network, Tohoku University, Sendai 980-8577, Japan.

[f]Department of Physics, Osaka University, Toyonaka, 560-0043, Japan

[g]Department of Physics, University of Tokyo, Hongo, Tokyo 113-0033, Japan

[h]Department of Physics, ETH Zurich, 8093 Zurich, Switzerland

[i]Electronics and Photonics Research Institute Advanced Industrial Science and Technology (AIST) Tsukuba, Ibaraki 305-8568, Japan

[j]CREST, Japan Science and Technology Agency, Saitama 332-0012, Japan

Abstract: Graphene with excellent 2D characters has been required to expand their potentials for meeting applicational demands in recent years. As one avenue, several approaches for fablicating 3D porous graphene network structures have been developed to realize multi-functional graphene materials with 2D graphene. Here we overview characteristics of 3D porous graphene for future electronic device applications along with physical insights into "2D to 3D graphene", where the characters of 2D graphene such as massless Dirac fermions are well preserved. The present review thus summarizes recent 3D porous graphene studies with a perspective for providing new and board graphene electronic device applications.

# 1. Introduction

Two-dimensional (2D) materials have been central to the developments of various kinds of potentiality as well as practical applications. Especially, graphene-related materials have become crucial research targets and have proposed fascinated applications on electrical transport[1–5], thermal transport[6], optical[7–12], plasmon[13–16], biochemical[17–20], filter/sensing[21–25] and energy[26–33] devices. Those devices successfully exploit the graphene characteristics. Specifically, electronic devices from 2D graphene sheets exhibit promising functionalities such as transistors[34–36], flexible and transparent electrodes[37] and displays[38]. These prototype devices have shown that graphene-based electrical devices can indeed be applied to practical applications. At the same time, however, it has been recognized that the performances in 2D graphene-based devices are sometimes much lower than those of conventional carbon devices, which implies that a single graphene sheet itself cannot cover various kinds of applications. Hence graphene materials are facing big challenges for further developments and expansion of applicational breadths.

Recently, three-dimensional (3D) porous graphene network architectures constructed from graphene sheets are receiving focused attentions for expanding graphene applications. Various kinds of 3D porous graphene materials have been created, which includes chemically exfoliation, sol-gel methods, template methods and chemical vapour deposition (CVD) for applications such as supercapacitors[39–50], biochemical applications[51–53] lithium batteries[54–64], electrocatalysts[65–74], photodetectors[75,76], sensing and filter devices[77–86], mechanical applications[87–90], water purification[91–93], transistors[94] and plasmonics[95]. These 3D-graphene based devices have demonstrated high performances beyond 2D-graphene devices. However, such 3D porous graphene devices often suffer from undesirable phenomena and characteristics caused by their 3D morphology itself. For example, uncontrollable connections of graphene layers may bring about electrical short circuits with leakage currents, preventing fine electrical device control. Moreover, 3D morphology, when non-porous, could hinder mass transport of ions and molecules required for chemical reactions and electrical charge transfers. This is natural, since physical properties, hence applications, of 3D porous graphene devices should strongly depend on the structural morphology and their electronic characters in which 2D graphene is modified. Here we envisage that 3D porous graphene devices for realizing various kinds of graphene based electronic devices are subject to requirements of bi-continuous, monolithic, and highly-crystalline structures with open porosity, which can then realize high electrical conductivities and mobilities, large surface areas, high mechanical strengths, high thermal conductivities and chemical stabilities with well-preserved 2D graphene natures of graphene sheets.

Indeed, one crucial point in 3D porous graphene studies is how 2D characters are preserved in 3D structures. In general, the 2D graphene characters such as massless Dirac fermions and excellent transport properties may be lost due to 3D structures with curved graphene sheets, where topological defects such as 5-7 defects are geometrically required. We have also to take care of disorder arising from discontinuous graphene flakes with frayed borders and edge defects, and distorted/disordered graphene grains created in fabrication processes. Therefore, it could be expected that 3D porous graphene-based electronic devices become more attractive if the 2D graphene characters are well preserved in 3D porous structures. Quite recently, it has been reported that a class of 3D porous graphene preserves 2D graphene characters associated with massless Dirac fermions with high electron conductivity and mobility.[96] By combining conventional graphene characters with 3D porous structures, the high crystalline 3D porous graphene with inter-connected open porous structures gives a new path for interesting physical properties and applications. The present review gives a perspective in current 3D graphene fabrication methods along with physical properties influenced by 3D morphologies for broadening graphene-based electronic device applications to overcome limitations beyond 2D graphene devices.

## 2. Synthesis of structure-controlled 3D porous graphene
### 2.1 Chemical treatment method

The self-assembly or sol-gel method is a wet chemical process of chemical exfoliation followed by an assembly of the exfoliated graphene nanosheets into open porous structures, with ultralow density and large surface area with and without other components, such as metal oxides (**Fig. 1**). Most commonly used processes are the modified Hummers' method for producing chemically exfoliated reduced graphene oxide (RGO) nanosheets with good dispersibility and processability, and standard sol-gel methods with freeze drying or supercritical drying techniques to keep the porous structures.[75,97-105] The highest specific surface area (3000 m$^2$g$^{-1}$) with ultra-light density (0.2 mg cm$^{-3}$) in the graphene materials is created by the sol-gel method.[106,107] Advantages of the two methods by wet chemical processes are easy fabrications of 3D architectures and applicability to mass production with desirable functionalization by doping of metal oxide nanoparticles (NPs) or other carbon materials such as carbon nanotube (CNT) for applications in supercapacitors, biosensors and electrocatalysts.

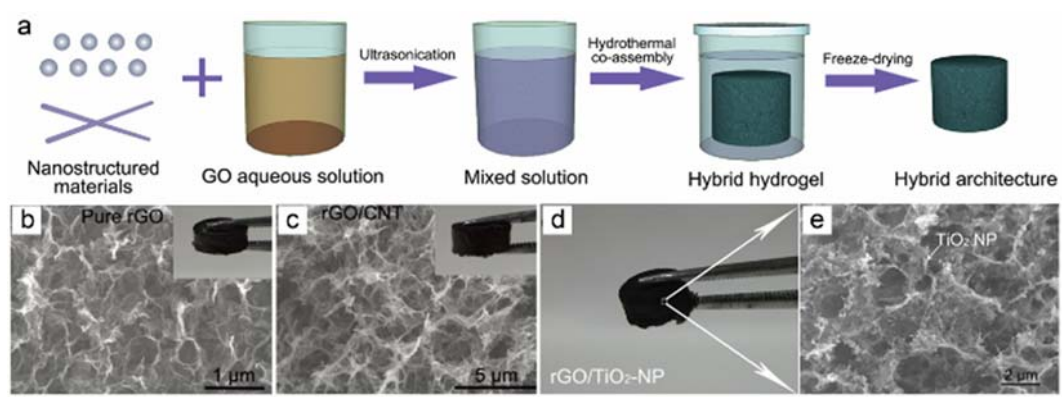

**Fig. 1** A universal approach for the synthesis of graphene hydrogel for 3D porous structures. (a) A schematic preparation of porous graphene hybrid architectures by a hydrothermal co-assembly process. (b) Pure RGO architecture, (c) rGO/CNTs architecture, (d-e) RGO/TiO$_2$-NP hybrid architecture. Reproduced from ref. 75 with permission.

**2.2 Hard template based method**

Controlled graphene structures have also been fabricated by using hard templates such as spheres, tubes and zeolites, where graphene replicates the templates structures. The templates are carefully selected in terms of thermal and chemical stability during precursor filling into the templates and annealing at high temperatures for graphitization. In the hard template method, both templates and filled precursors are annealed mostly over 1000 ℃, and pure graphitic materials are obtained by chemical etching of the templates. The advantages of template methods can achieve a fine design of porous and periodic structures with changeable template sizes/shapes and configurations. For example, SiO$_2$[108−110], Al$_2$O$_3$[111,112], CaCO$_3$[113] or polystyrene[114−116] are used to obtain the holy-like 3D porous graphitic structures for supercapacitors and other energy devices (**Fig. 2**).

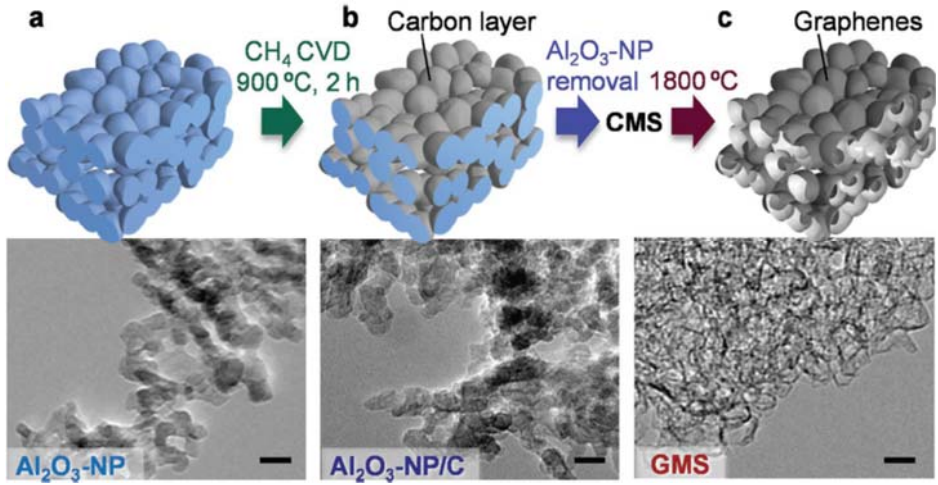

**Fig. 2** Schematic synthesis of graphene mesosponges. $Al_2O_3$-nanoparticles was employed as a nanosized CVD substrate for graphitization of carbon mesosponge for a high-temperature treatment. Reproduced from ref. 111 with permission.

**2.3 Microporous template based CVD method**

The bottom-up CVD is known as one of the best fabrication methods for obtaining largest, bicontinuous and high quality defect-free graphene with large grain sizes.[117–120] The graphene growth mechanism during CVD processes can be explained with the surface catalytic reaction on Cu surface with carbon segregation and precipitation on Ni surface at temperatures around 400-1000 ℃. Cheng and coworkers first demonstrated 3D microporous fabrication with the CVD method by employing Ni foam, which brought an important breakthrough for bi-continuous and 3D microporous graphene.[121] This concept is successfully applied to other metal templates such as metal complex powders, metal nanoparticles and porous metals in place of Ni foam to achieve porous graphene.[123–127] There, microporous structures with length scale of 10-500 μm preserve 2D graphene characters, which is why the graphene foams created from Ni foam have been widely employed for electrical-device and energy-related applications. However, these graphene foams exhibit relatively poor electrical transport properties for devices, due to the nature of Ni foam with large surface roughness and the surface connection that is not smooth.

Moreover, the low curvature (with over 10 μm diameter pore size) in the 3D porous graphene structures should exerts little influences on their physical properties, but the relationship between 3D porous graphene properties and curvatures have not been discussed in detail.

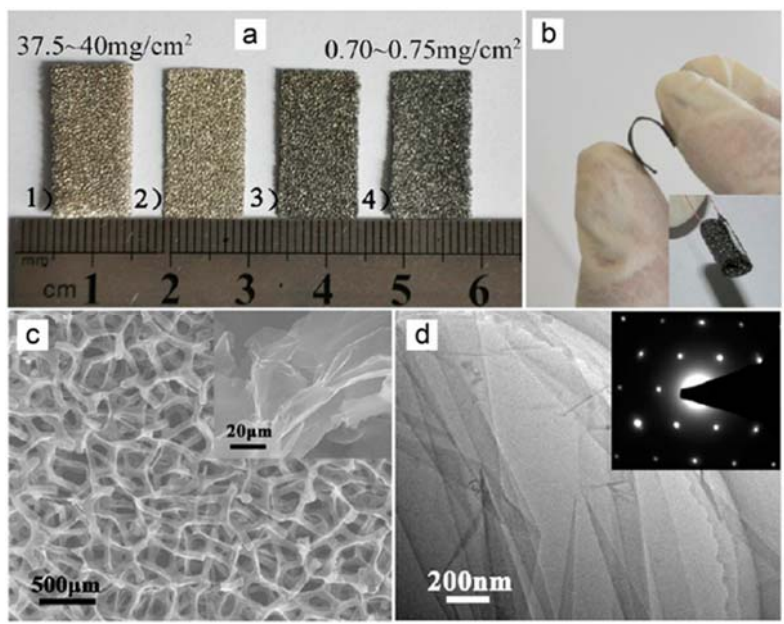

**Fig. 3** Microporous template based 3D porous graphene by using Ni foam. (a) Digital photographs of (1) Ni foams and (2) pressed Ni foam, and graphene-coated pressed Ni foams (3) before and (4) after removal of the Ni foam. (b) A freestanding and flexible 3D porous graphene network prepared from a pressed Ni foam. Inset shows curled 3D porous graphene networks. (c) SEM image of a 3D porous graphene network after removal of Ni foam. Inset shows a high-magnification image. (d) TEM image of a graphene sheet. Inset shows a SAED pattern of the corresponding graphene nanosheets. Reproduced from ref. 120 with permission. Copyright 2013 American Chemical Society.

**2.4 Nanoporous metal based CVD method**

More recently, the Ni foam has been replaced by nanoporous metals such as Ni[128,129] or Cu[130,131] with initial pore size of about 10 nm for obtaining 3D nanoporous graphene with smaller nano-sized porous structures. The advantage of such metal templates is not only nano-sized porous structures but also their thermally reconstructed very smooth surfaces and the unique structures such as a triply-periodic minimal surface which comprises 55% gyroid surface and 30% Schwarz's D-surface.[132,133] By using the nanoporous metals as CVD templates, it is reported that the obtained high quality, bi-continuous and open 3D nanoporous graphene preserves the massless Dirac fermion character of 2D graphene with tunable porosity and high surface area, as displayed in **Fig. 4**.[96] The morphology of the nanoporous graphene traces the nanoporous Ni template,[128] and the freestanding, bi-continuous and monolithic nanoporous graphene is isolated by chemical etching of the Ni template with weak acid treatments. Moreover, this CVD approach can accommodate chemical dopants into graphene lattices on a 3D graphene sheet with chemical precursors such as pyridine, thiophine and phosphorus as carbon and chemical dopant sources. The chemically doped 3D nanoporous graphene has chemical and physical properties successfully tuned, and brings chemical activities enhanced by defects induced by the chemical dopants into chemically inert graphene as metal-free electrode catalysts for applications such as fuel cell, hydrogen evolution and battery.[58,65,70,76,92]

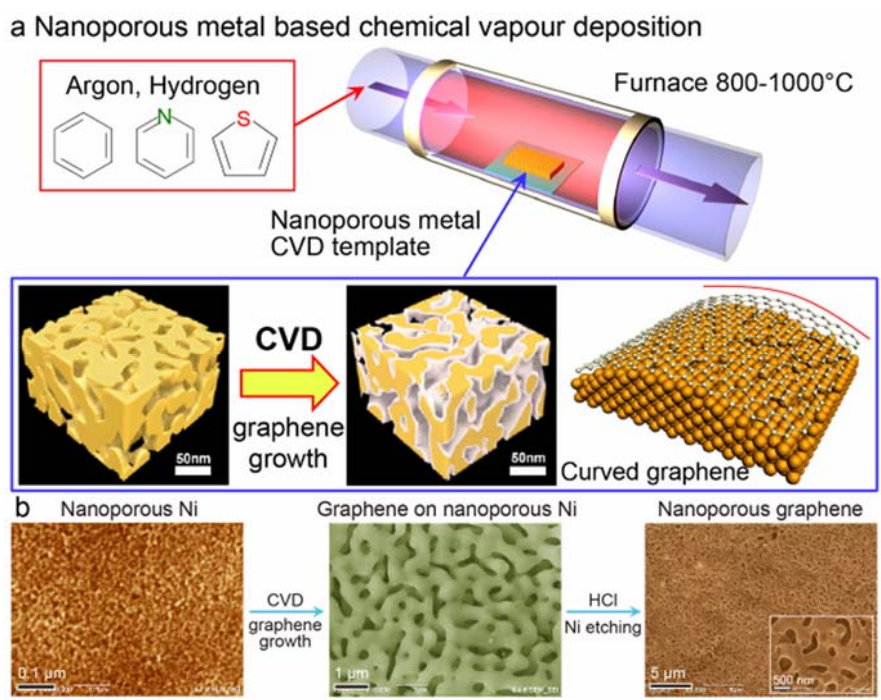

**Fig. 4** Nanoporous template based 3D nanoporous graphene by using nanoporous Ni. (a) A schematic synthesis of nanoporous graphene. Nanoporous Ni was employed as a nanopore CVD substrate for graphene growth with different kinds of precursors. (b) Preparation of nanoporous graphene by nanoporous Ni templates.

## 3. Hallmarks of high quality 3D nanoporous graphene

Constructing 3D nanoporous graphene with millimeter-sized architecture from an ideal single 2D sheet without compromising its remarkable electrical, optical, and thermal properties is currently a great challenge for overcoming limitations in integrating single graphene flakes into 3D porous graphene devices. Inter-connected and bicontinuous 3D nanoporous graphene without frayed borders/edges and with low density of crystalline defects very similar to those of suspended graphene layers have been investigated by combining the state-of-the-art spectromicroscopy and imaging techniques of Raman spectroscopy, TEM observation and photoemission spectroscopy.[96,134] The high spatial resolution brings to light the inter-relationship between the topology and the morphology in the highly curved regions where $sp^2$ carbon bonds are hybridized with $sp^3$ states. Moreover, the space- and angle-resolved analysis of the electronic states via the photoelectron spectroscopy associates different hybridization states and electronic spectral density of state (DOS) with the different (flat and curved) spatial regions in the 3D nanoporous graphene.[96,134]

### 3.1 Morphology and structures of 3D nanoporous graphene

Now, a 3D nanoporous graphene with high interconnectivity, low defect density, and tunable nanopore sizes is obtained by the nanoporous Ni-based CVD method.[96] The microstructure of 3D graphene shows the intricate 3D nanoporous morphology with concave and convex curvatures and nanopores (**Fig. 5**). The selected area electron diffraction reveals that the 3D nanoporous graphene has multiple orientations, which is associated with random distributions of the interconnected highly crystalline graphene sheets in three dimensions. The atomic structure of the 3D nanoporous graphene shows a perfect hexagonal structure on the flat region along with topological defects such as a 5–7 defect in large curvature regions. These atomic-scale defects are geometrically required for realizing 3D nanoporous configurations.

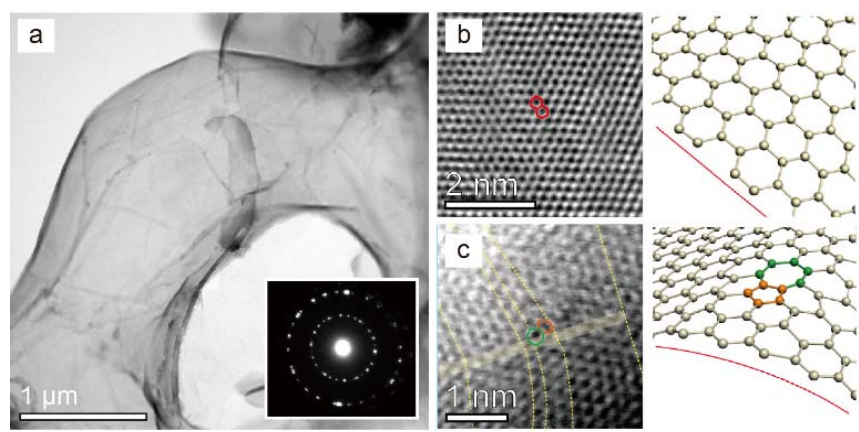

**Fig. 5** TEM image of 3D nanoporous graphene. (a) Typical low-magnification image with the selected area electron diffraction pattern. (b) High resolution TEM image on a flat region in the 3D nanoporous graphene with the corresponding atomic model. (c) The same for a curved region with the corresponding atomic model with a 5-7 defect.

A selected micro-Raman spectrum in **Fig. 6** on the 3D nanoporous graphene demonstrates the presence of a high-quality graphene constituted by a majority of interconnected single layers due to the low D band intensity and the high 2D band intensity.[134] The histogram of the $I_{2D}/I_G$ ratio (**Fig. 6a-b**) indicates a high average graphene quality ($I_{2D}/I_G$ ratio = 2.6, σ = 0.6) accommodating a variety of morphological configurations/ orientations. This high value of $I_{2D}/I_G$ is consistent with the presence of one to two layers of planar graphene. A quality hallmark of the 3D nanoporous graphene is seen from a $I_{2D}/I_G$ intensity ratio mapping in **Fig. 6c**: Sub-nanometer flat areas are clearly identified from $I_{2D}/I_G$, in agreement with TEM images (**Fig. 5**). The $I_D/I_G$ mapping reflecting the defect distribution density is inversely proportional to the $I_{2D}/I_G$ image in terms of brighter versus darker regions (**Fig. 6d**). This mapping shows that a relatively high defect concentration exists in the tubular-shaped and highly-curved regions, where the defects such as 5–7 defects should be geometrically required to form curved graphene as discussed in **Fig. 5**.

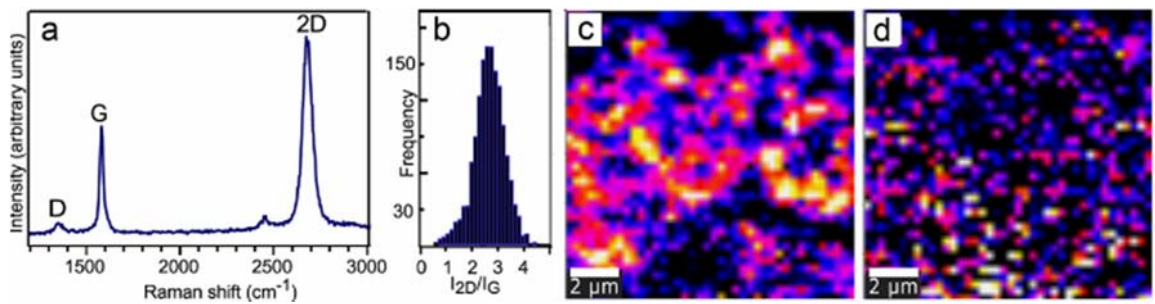

**Fig. 6** Raman spectroscopy of 3D nanoporous graphene. (a-b) Micro-Raman spectrum taken on a 500 nm-diameter spot with the distribution of the 2D/G integral intensity ratio taken on a 10 × 10 μm². Spatially resolved micro-Raman maps of (c) 2D/G and (d) D/G intensity ratios; 12 × 12 μm² images formed by 300 × 300 nm² pixels. Reproduced from ref. 134 with permission. Copyright 2017 American Chemical Society.

**3.2 Electronic structure in 3D nanoporous graphene**

The spatially resolved C 1s core-level mapping obtained by X-ray photoelectron spectroscopy (XPS) identifies two components of $sp^2$-like (284.4 eV) and $sp^3$-like (285.1 eV) states (**Fig. 7**). The relative intensity ratio of the two components varies in the mapping, depending on the topology of the flat or curved 3D nanoporous graphene regions. The tubular-shaped ligament structures are also reflected in the $sp^2$-like mapping. The intensity spatial mappings demonstrate that flat regions are mainly dominated by the $sp^2$ component and the border areas of the elongated tubular and highly curved regions are prevailingly dominated by the $sp^3$-like component. Moreover, the tubular and curved regions could be induced partially warped $sp^2$ bonding and rehybridize toward $sp^3$-like configurations, generating a high binding energy weight in the spectra. Furthermore, C 1s peaks related with edge defects are not detected, consistent with the bicontinuous topology in (**Fig. 5**) and the low D band intensity in the Raman spectra (**Fig. 6**).

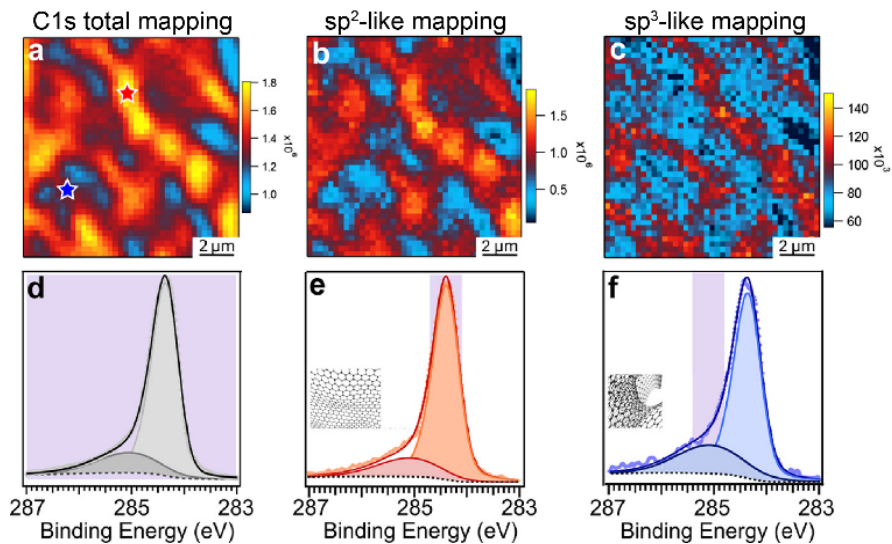

**Fig. 7**. XPS results for the 3D nanoporous graphene. (a) Spatially resolved C 1s core-level mapping of the sample on 12 × 12 μm$^2$ area, constituted by 300 × 300 nm$^2$ pixels, taken with 350 eV photon energy. (b-c) The sp$^2$-like and sp$^3$-like mappings. The mapping was obtained from (e) intensities of sp$^2$-like peak (shaded) and (f) intensities of sp$^3$-like peak (shaded). (d) The C 1s core-level spectrum spatially averaged over the mapping. (e) The spectra with the dominant sp$^2$-like component (red star in the total mapping in (a)) associated with the flatter areas. The peaks highlighted in red were assigned to be sp$^2$-like and sp$^3$-like. (f) The spectra with more intense sp$^3$-like components (blue star in (a)) associated with the highly bent/wrinkled regions highlight their complementary spatial distribution. The peaks highlighted in blue were assigned to be sp$^2$-like and sp$^3$-like. Color scales of maps are normalized to the maximum intensity of the corresponding peak. Reproduced from ref. 134 with permission. Copyright 2017 American Chemical Society.

If graphene in 3D nanoporous structures retains a massless Dirac fermion character (**Fig. 8(a)**) with high electron mobility, this is expected to serve as a post-silicon electrical device in near future. This will also facilitate commercializing graphene devices. Recently, a 3D nanoporous structure comprising monolayer graphene with Dirac fermions preserved has been synthesized, and the electronic structure of the nanoporous graphene as investigated with photoemission spectroscopy showed a Dirac-cone type linear electronic DOS around the Fermi energy that resembles that of 2D graphene (**Fig. 8(b-c)**). However, contrary to 2D graphene which has an anisotropic angle-dependence for photoelectrons emitted from well-ordered samples in angle-resolved PES, the 3D nanoporous graphene exhibits no significant anisotropy in ARPES (**Fig. 8(d)**).[96] If we recall the major sp$^2$-like states in the 3D graphene (**Fig. 7**), this indicates that the 3D nanoporous graphene preserves a graphene-like electronic structure, i.e., the massless Dirac fermions, and the electron scattering on the 3D periodic structures, that can in general affect physical properties, with e.g. a mass-gap opening for atomic-scale 3D periodicities[135], may be suppressed in this system.[96,134] The 3D nanoporous graphene overcomes the limitations in practical applications of small-sized individual graphene sheets for 3D devices constructed from a single 2D graphene sheet.

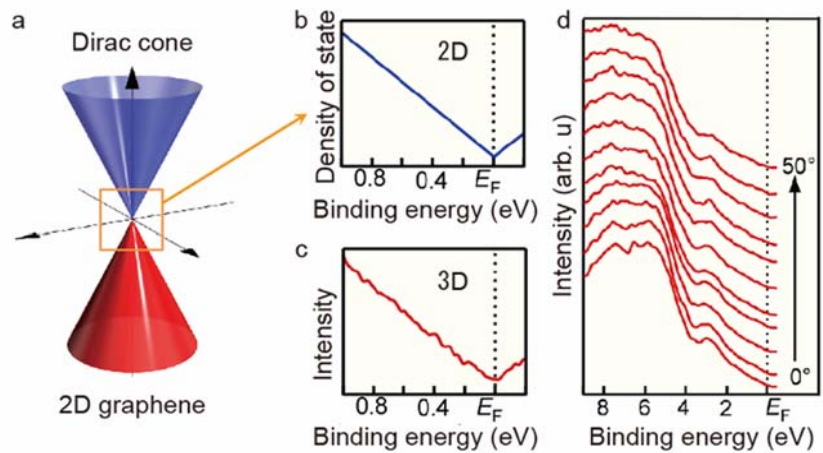

**Fig. 8**. PES results for 3D nanoporous graphene. (a) Conventional 2D graphene Dirac cone dispersion. (b) and (c) Electronic DOS around the Dirac point in 2D and 3D nanoporous graphene, respectively. The DOS was obtained by dividing the PES spectrum with the Fermi-Dirac function convoluted with the instrumental energy resolution. (d) Angle-resolved PES spectra in a wider energy region for 3D nanoporous graphene.

**4. Transport properties in 3D porous graphene**

The electric transport properties of 3D porous graphene materials reviewed in **sections 2-3** play an important role in their electronic device performances, and have been investigated for obtaining a deep understanding for potential 3D device applications. The performance turns out to strongly depend on the synthesis and fabrication methods (**Fig. 9**), whose electrical conductivity values are obtained from the pore size and conductivity in the reported literatures.[136] The chemically treated graphene materials obtained by reducing a graphene oxide with sol-gel methods demonstrate a wide range of pore sizes from 4 nm to 2 μm with electrical conductivity ranging 0.01−10$^4$ S/m. This may be compared with the graphene foams fabricated with Ni foam with abundant micro-sized pores that have electrical conductivity ranging 10−10$^3$ S/m.

The nanoporous metal-templated graphene reviewed in **section 3** shows high conductivity ranging 10$^3$−10$^4$ S/m with 0.1−1 μm pores. However, the corresponding electron mobility, in a range of 100−500 cm$^2$/Vs at various temperatures, turns out to be nominally 10−50 times lower than the mobility of 2D CVD graphene (**Fig. 9(b)**). However, it should be noted that the mobility was estimated by taking the nanoporous graphene sample as a uniform 3D metal, i.e., the electron mobility is simply calculated from the 3D conductivities normalized by the 3D sample dimensions (thickness, etc). Since electron trajectories should be complex on curved graphene surface in nanoporous structures, the mobility there cannot be compared in such a simple manner to that of a flat 2D graphene. However, it is possible to estimate the actual mobility on local graphene surfaces in 3D nanoporous structure if one turn to the magnetic transport measurement as described in the following.

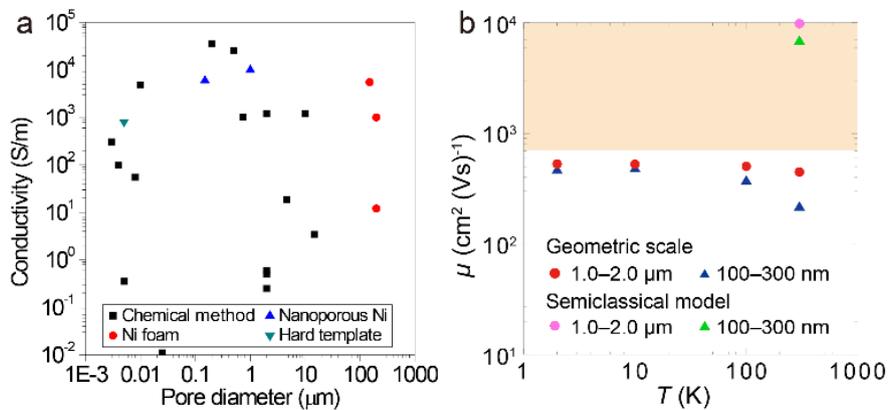

**Fig. 9** (a) Electrical conductivity vs pore size for various 3Dporous graphene samples reported in the literatures. (b) Electron mobility vs temperature in 3D nanoporous graphene with various porous sizes. The orange region presents the mobility of conventional 2D CVD graphene.

The electrical transport properties of the porous materials (as opposed to ordinary metals) demonstrate an interesting phenomenon related with their unique 3D periodic structures. [132,133] Especially, estimation of electron mobility should be greatly influenced by the electron pathways in nanoporous structures (**Fig. 10**). Here is one example for the difference between estimations for a graphene nanoporous structure and the 3D bulk sample in a semiclassical model. **Figure 10(a)** schematically displays a part of nanoporous graphene ligament placed in a uniform magnetic field ***B***. During traveling along a pathway in a monolithic nanoporous graphene with the relative angle between the magnetic field and the curved graphene surface changing from one place to another, an electron feels spatially varying magnetic fields ***B**$_{eff}$* (= $B\cos\theta$) for the component normal to the surface, where $\theta$ is the angle between ***B*** and the surface normal. Thus an electron travelling on a graphene labyrinth experiences a non-uniform magnetic field, so that the cyclotron motion, dominated by the normal component of the magnetic field, should be affected. This means that the total conductivity is expressed as an appropriate average of the 2D conductivity over a magnetic field distribution when the inhomogeneity is spatially slowly-varying enough. In that case we can take the semiclassical approximation, in which the conductivity tensor is given as

$$\sigma_{ij}(B) = \frac{S_{tot}}{V} \int_0^{\frac{\pi}{2}} \sigma_{ij}^{2D}(B\cos\theta)\sin\theta d\theta$$

where $\theta$ is again the angle between the surface normal and the magnetic field, $S_{tot}$ and $V$ are the total graphene area and the total volume of the system, respectively.[94] The Hall resistivity obtained from this expression is plotted in **Fig. 10(c)** as compared with the experimental result. Here the theoretical result predicts that the Hall resistivity is not perfectly linear in $B$ but has a slight kink structure as seen in **Fig. 10(c)**. Accordingly, the second derivative, for a more detailed characterization of the Hall resistivity curve, exhibits a peak at $\mu B \sim 0.7$ where $\mu$ is the local mobility of 2D graphene in the 3D nanoporous structure. Conversely, this implies the local mobility $\mu$ can be estimated from the peak position of the second derivative of the Hall resistivity against $B$. The theory reproduces the experimental curve qualitatively well, and the mobility $\mu$ estimated from the peak position is 5000−10000 cm$^2$/Vs, which is comparable to the electron mobility for 2D CVD graphene (**Fig. 9(b)**). Such electron trajectories in the complex 3D porous structures should be considered in their transport properties for a deep understanding of

3D porous graphene based electronic devices.

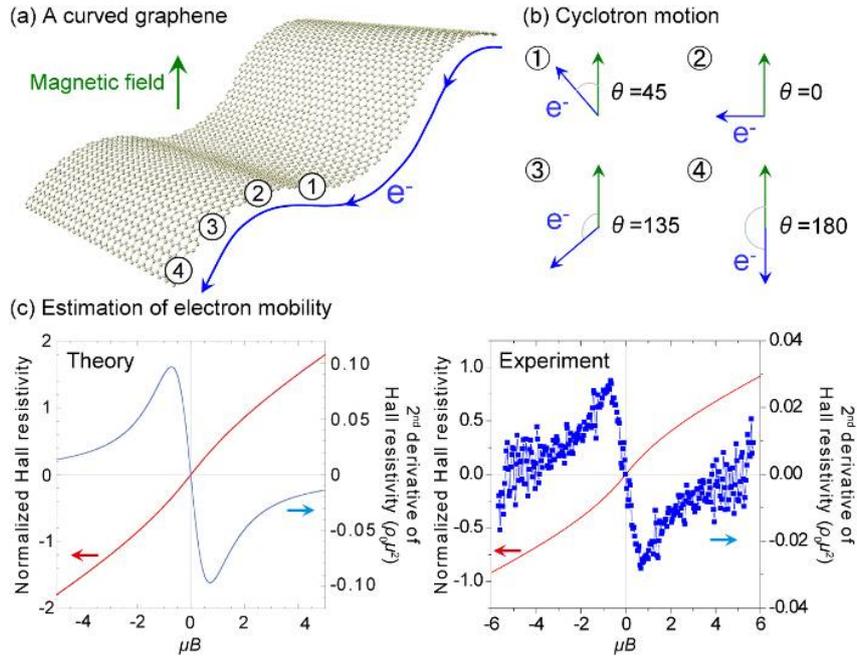

**Fig. 10** (a) Propagation of an electron on a curved graphene as a part of the nanoporous structure in a magnetic field ***B***. (b) Angle between ***B*** and the local surface normal at various positions marked in (a). Theoretical (c) and experimental (d) results for the Hall resistivity (red curves) and its second derivative (blue) against $\mu B$, where $\mu$ is the carrier mobility.

**5. 3D nanoporous graphene transistor**

   Free-standing, large-scale 3D nanoporous graphene with 2D electronic properties, such as high electrical conductivity and electron mobility as one successful electronic device example, holds great promise for applications, especially as a 3D nanoporous graphene transistor[94,137]. A high performance 3D nanoporous graphene transistor has actually been demonstrated using the electric double layer (EDL) capacitance[35]. The vast surface area in bicontinuous and open nanoporous structures of graphene should indeed give significant advantages for both transistor channels and gate electrodes. Employing the nanoporous graphene as a transistor channel and a gate electrode to maintain sufficient electrical capacitance as shown in **Fig. 11(a)**, a carrier density of a wide-area graphene sheet integrated into a 3D nanoporous structure was uniformly controlled by the electric field effects. Clear ambipolar carrier transport features were observed in the longitudinal and transverse conductance as well as in the quantum capacitance that reflects the electronic structure of the massless Dirac fermion (**Fig. 11(b)**). The bicontinuous and open nanoporous structure in electric fields is also clarified from the magnetotransport scaling as previously discussed in **section 4**. Importantly, both the conductance and the capacitance are 100–1000 times higher than those in the 2D graphene EDL transistor (**Fig. 11(c)**)[35]. Thus we can expect the 3D nanoporous graphene EDL transistor will open a new horizon for applications in highly responsive electronic devices as well as for tuning electronic properties of Dirac fermions with the 3D nanoporous structure.

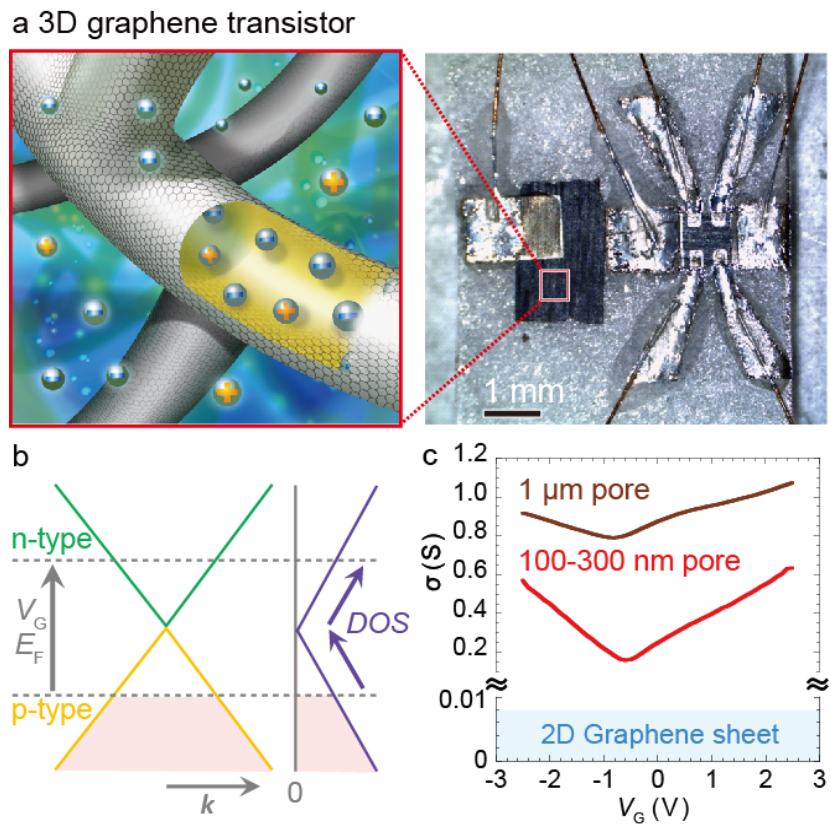

**Fig. 11** 3D nanoporous graphene transistor. (a) Schematic carrier doping and an optical image of a real device. (b) A schematic electronic dispersion for the 3D nanoporous graphene at various positions of the gate voltage $V_G$. (c) Dependence of the conductivity $\sigma$ on $V_G$ for the 3D nanoporous graphene samples with pore sizes of 100-300 nm and 1 μm, respectively. For comparison, the shaded region represents the 2D graphene EDLT in the literature[35].

## 6. Conclusions

In this article, we have reviewed the electronic structure and controllability of 3D porous graphene network and its hybrid architectures. Especially, 3D nanoporous graphene created from the nanoporous Ni based CVD method preserves a single and high-quality graphene sheet characters hallmarked by investigations of Raman spectroscopy, TEM observation and photoemission spectroscopy. The preservation of 2D graphene characters and understanding of physical properties of 3D porous graphene are the core of material science and physics for expanding 3D porous graphene based electronic applications. Using such high quality 3D porous graphene, we envisage they have various potential and promising device applications such as supercapacitors, lithium batteries, electrode catalysts (hydrogen evolution reaction, oxygen reduction reaction and oxygen evolution reactions) by utilizing renewable energy sources, solar light absorption materials, transistor, photodetector, sensing devices as metal-free and/or eco-friendly materials. One message is that the graphene morphology and fabrication methods should be properly chosen according to the usage and applications. Even when the morphology of a single and perfect 2D graphene sheet demonstrates excellent characters, atomically-thin sheets are not always fit for various kinds of applications, and graphene materials may harbour wider possibilities for practical and industrial applications when 2D graphene sheets are converted to 3D porous architectures, so that the 3D porous graphene reviewed here has a great potential towards future applications. However, proper and deeper understanding of their physical properties, especially electrical transport properties, is desirable with theoretical approaches[138]. As we have stressed, the electron pathways in 3D porous structures should be quite complicated and much longer than direct distances. Therefore, we shall have to further elaborate these for capturing 3D porous graphene characters and functionalities for developments of graphene based electronic devices.

Given a wide variety of carbon structures, the unique properties of 3D porous graphene architectures may be understood by combining the features of graphite, graphene, carbon nanotube, nanoribbon, nanographene and fullerene, which means that the 3D porous graphene materials incorporates some of these functions as schematically illustrated in **Figure 12**. Therefore, it could be expected that 3D porous graphene devices will bring new insights and innovations to overcome the limitation of integrating graphene in 3D devices by fusion of advantages of carbon family for opening a new route for a plethora of 3D porous graphene applications.

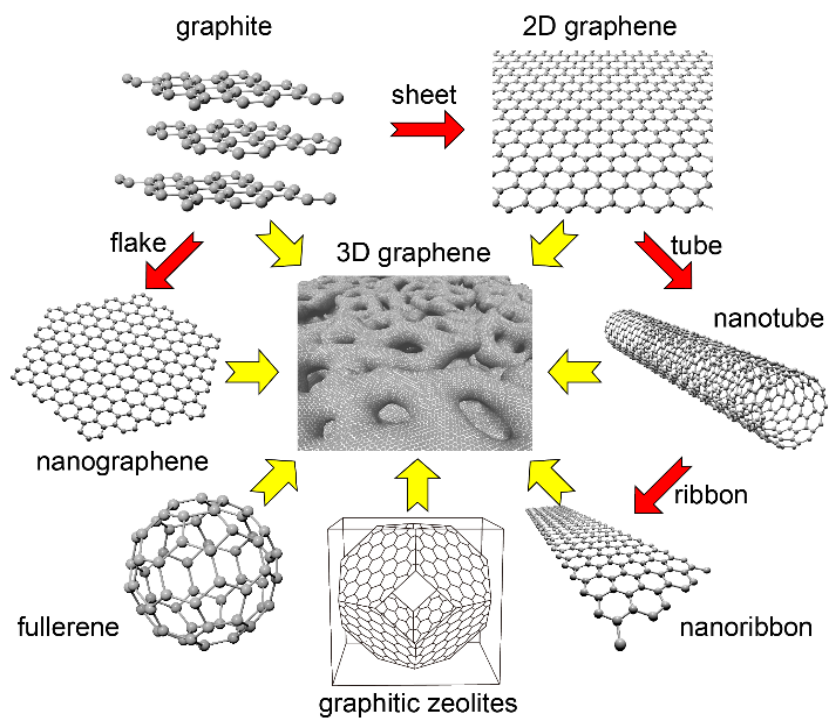

**Fig. 12** 3D porous graphene as a composite of various carbon families.

**Conflicts of interest**

There are no conflicts to declare.

**Acknowledgements**

We would like to thank Prof. Toshiaki Enoki and Prof. Yoji Koike for valuable discussion and Mr. Masahiko Izumi for 3D graphene illustration. This work was sponsored by JST-CREST "Phase Interface Science for Highly Efficient Energy Utilization"; the fusion research funds of "World Premier International (WPI) Research Center Initiative for Atoms, Molecules and Materials", JSPS KAKENHI Grant Numbers JP23224010, JP24740216, JP24656028, JP24740193, JP25107001, JP25107003, JP25107005, JP25249108, JP26107504, JP26247064, JP26289294, JP15H05473, JP17K05496, JP17K14074. This work was partly performed at High Field Laboratory for Superconducting Materials, Institute for Materials Research, Tohoku University (Project

No. 17H0201).